\documentclass[twocolumn, twocolappendix]{aastex63}

\usepackage{xspace}
\usepackage{newtxmath,newtxtext}
\usepackage{CJK}
\usepackage{dsfont}

\newcommand{\logg}{\ensuremath{\log g}\xspace}
\newcommand{\feh}{\ensuremath{[\mathrm{Fe}/\mathrm{H}]}\xspace}
\newcommand{\alphafe}{\ensuremath{[\alpha/\mathrm{Fe}]}\xspace}
\newcommand{\mgfe}{\ensuremath{[\mathrm{Mg}/\mathrm{Fe}]}\xspace}
\newcommand{\alfe}{\ensuremath{[\mathrm{Al}/\mathrm{Fe}]}\xspace}
\newcommand{\nife}{\ensuremath{[\mathrm{Ni}/\mathrm{Fe}]}\xspace}
\newcommand{\gaia}{\textit{Gaia}\xspace}

\newcommand{\jphi}{\ensuremath{J_\phi}\xspace}
\newcommand{\jr}{\ensuremath{J_R}\xspace}
\newcommand{\jphiunits}{\ensuremath{\mathrm{Mpc}\,\mathrm{km}\,\mathrm{s}^{-1}}\xspace}
\newcommand{\jrunits}{\ensuremath{[\mathrm{kpc}\,\mathrm{km}\,\mathrm{s}^{-1}]^{1/2}}\xspace}

\newcommand{\kms}{\ensuremath{\textrm{km}\,\textrm{s}^{-1}}\xspace}

\accepted{for publication by ApJL}

\shorttitle{The Nyx stream in GALAH}
\shortauthors{Zucker et al.}
\graphicspath{{./}{figures/}}

\begin{document}
\begin{CJK*}{UTF8}{gbsn}

\title{The GALAH Survey: No chemical evidence of an extragalactic origin for the Nyx stream}

\correspondingauthor{Daniel~B.~Zucker}
\email{daniel.zucker@mq.edu.au}
\author[0000-0003-1124-8477]{Daniel~B.~Zucker}
\affiliation{Department of Physics and Astronomy, Macquarie University, Sydney, NSW 2109, Australia}
\affiliation{Macquarie University Research Centre for Astronomy, Astrophysics \& Astrophotonics, Sydney, NSW 2109, Australia}
\affiliation{Centre of Excellence for Astrophysics in Three Dimensions (ASTRO-3D), Australia}

\author[0000-0002-8165-2507]{Jeffrey~D.~Simpson}
\affiliation{School of Physics, UNSW Sydney, NSW 2052, Australia}
\affiliation{Centre of Excellence for Astrophysics in Three Dimensions (ASTRO-3D), Australia}

\author[0000-0002-3430-4163]{Sarah~L.~Martell}
\affiliation{School of Physics, UNSW Sydney, NSW 2052, Australia}
\affiliation{Centre of Excellence for Astrophysics in Three Dimensions (ASTRO-3D), Australia}

\author[0000-0003-3081-9319]{Geraint~F.~Lewis}
\affiliation{Sydney Institute for Astronomy, School of Physics, A28, The University of Sydney, NSW 2006, Australia}

\author[0000-0003-0174-0564]{Andrew~R.~Casey}
\affiliation{School of Physics and Astronomy, Monash University, Australia}
\affiliation{Centre of Excellence for Astrophysics in Three Dimensions (ASTRO-3D), Australia}

\author[0000-0001-5082-9536]{Yuan-Sen Ting  (丁源森)}
\affiliation{Institute for Advanced Study, Princeton, NJ 08540, USA}
\affiliation{Department of Astrophysical Sciences, Princeton University, Princeton, NJ 08544, USA}
\affiliation{Observatories of the Carnegie Institution of Washington, 813 Santa Barbara Street, Pasadena, CA 91101, USA}
\affiliation{Research School of Astronomy \& Astrophysics, Australian National University, ACT 2611, Australia}

\author[0000-0002-1160-7970]{Jonathan Horner}
\affiliation{Centre for Astrophysics, University of Southern Queensland, Toowoomba, QLD 4350, Australia}

\author[0000-0001-5344-8069]{Thomas Nordlander}
\affiliation{Research School of Astronomy \& Astrophysics, Australian National University, ACT 2611, Australia}
\affiliation{Centre of Excellence for Astrophysics in Three Dimensions (ASTRO-3D), Australia}

\author[0000-0002-4013-1799]{Rosemary~F.~G.~Wyse}
\affiliation{1Department of Physics \& Astronomy, Johns Hopkins University, Baltimore, MD 21218, USA}

\author[0000-0002-2325-8763]{Toma\v{z}~Zwitter}
\affiliation{Faculty of Mathematics and Physics, University of Ljubljana, Jadranska 19, 1000 Ljubljana, Slovenia}

\author[0000-0001-7516-4016]{Joss~Bland-Hawthorn}
\affiliation{Sydney Institute for Astronomy, School of Physics, A28, The University of Sydney, NSW 2006, Australia}
\affiliation{Centre of Excellence for Astrophysics in Three Dimensions (ASTRO-3D), Australia}

\author[0000-0002-4031-8553]{Sven~Buder}
\affiliation{Research School of Astronomy \& Astrophysics, Australian National University, ACT 2611, Australia}
\affiliation{Centre of Excellence for Astrophysics in Three Dimensions (ASTRO-3D), Australia}

\author[0000-0002-5804-3682]{Martin~Asplund}
\affiliation{Max Planck Institute for Astrophysics, Karl-Schwarzschild-Str. 1, D-85741 Garching, Germany}

\author[0000-0001-7362-1682]{Gayandhi~M.~De~Silva}
\affiliation{Australian Astronomical Optics, Faculty of Science and Engineering, Macquarie University, Macquarie Park, NSW 2113, Australia}
\affiliation{Macquarie University Research Centre for Astronomy, Astrophysics \& Astrophotonics, Sydney, NSW 2109, Australia}

\author[0000-0002-2662-3762]{Valentina~{D'Orazi}}
\affiliation{Istituto Nazionale di Astrofisica, Osservatorio Astronomico di Padova, vicolo dell'Osservatorio 5, 35122, Padova, Italy}

\author[0000-0001-6280-1207]{Ken~C.~Freeman}
\affiliation{Research School of Astronomy \& Astrophysics, Australian National University, ACT 2611, Australia}

\author[0000-0001-7294-9766]{Michael~R.~Hayden}
\affiliation{Sydney Institute for Astronomy, School of Physics, A28, The University of Sydney, NSW 2006, Australia}
\affiliation{Centre of Excellence for Astrophysics in Three Dimensions (ASTRO-3D), Australia}

\author{Janez~Kos}
\affiliation{Faculty of Mathematics and Physics, University of Ljubljana, Jadranska 19, 1000 Ljubljana, Slovenia}

\author{Jane~Lin}
\affiliation{Research School of Astronomy \& Astrophysics, Australian National University, ACT 2611, Australia}
\affiliation{Centre of Excellence for Astrophysics in Three Dimensions (ASTRO-3D), Australia}

\author[0000-0002-8892-2573]{Karin~Lind}
\affiliation{Department of Astronomy, Stockholm University, AlbaNova University Centre, SE-106 91 Stockholm, Sweden}

\author[0000-0003-0110-0540]{Katharine J. Schlesinger}
\affiliation{Research School of Astronomy \& Astrophysics, Australian National University, ACT 2611, Australia}

\author[0000-0002-0920-809X]{Sanjib~Sharma}
\affiliation{Sydney Institute for Astronomy, School of Physics, A28, The University of Sydney, NSW 2006, Australia}
\affiliation{Centre of Excellence for Astrophysics in Three Dimensions (ASTRO-3D), Australia}

\author[0000-0002-4879-3519]{Dennis Stello}
\affiliation{School of Physics, UNSW Sydney, NSW 2052, Australia}
\affiliation{Centre of Excellence for Astrophysics in Three Dimensions (ASTRO-3D), Australia}

\begin{abstract}
The results from the ESA \textit{Gaia} astrometric mission and deep photometric surveys have revolutionized our knowledge of the Milky Way. There are many ongoing efforts to search these data for stellar substructure to find evidence of individual accretion events that built up the Milky Way and its halo. One of these newly identified features, called Nyx, was announced as an accreted stellar stream traveling in the plane of the disk. Using a combination of elemental abundances and stellar parameters from the GALAH and APOGEE surveys, we find that the abundances of the highest likelihood Nyx members are entirely consistent with membership of the thick disk, and inconsistent with a dwarf galaxy origin. We conclude that the postulated Nyx stream is most probably a high-velocity component of the Milky Way's thick disk. With the growing availability of large data sets including kinematics, stellar parameters, and detailed abundances, the probability of detecting chance associations increases, and hence new searches for substructure require confirmation across as many data dimensions as possible. 
\end{abstract}


\section{Introduction} \label{sec:intro}

In the $\Lambda$CDM paradigm for the formation and growth of galaxies and large-scale structures in the Universe, the growth of large galaxies like the Milky Way happens as a result of mergers with smaller bodies. 
As a smaller galaxy is accreted by the Milky Way, its stars will be tidally stripped into long tails, which can remain spatially coherent for many orbits because of the long dynamical time in the Galactic halo, and will retain their kinematic association for longer \citep[e.g.,][]{Freeman2002,Bullock2005,Johnston2008}. 
As a consequence, a history of accretion events will lead to an 
accumulation of stellar streams and tidal debris throughout the Milky Way; these structures serve as a fossil record of the events that created them, with examples present even in the Solar neighborhood \citep[e.g.,][]{Helmi1999}. Such 
structures provide opportunities to explore the gravitational potential of the Galaxy and the presence of substructure in the dark matter distribution, 
as well as the properties of the systems that have contributed to the growth of the Milky Way.

The \gaia mission \citep{GaiaCollaboration2016,GaiaCollaboration2018b} is revolutionizing this work, providing accurate and precise spatial and kinematic information for a huge sample of stars in the Galaxy. These data can be combined with the results from large spectroscopic surveys to identify and investigate streams and other stellar substructures of the Galaxy.

One of the major results from the Second Data Release of the \gaia mission \cite[\gaia DR2;][]{GaiaCollaboration2016,GaiaCollaboration2018b} was that a significant proportion of the halo stars near the Sun appears to have been accreted some 9 Gyr ago from a single dwarf galaxy, dubbed ``Gaia-Enceladus'' by \citet{Helmi2018}, and independently confirmed \citep[including][among others]{Myeong2018b,Myeong2018a,Haywood2018,Fattahi2019, Belokurov2018}. 

Spurred by this dramatic discovery, researchers have continued to sift through \gaia DR2 for evidence of more accretion events. Recently, one group searching for structures of accreted stars used a deep neural network classifier on the subset of stars for which \gaia provided astrometry and radial velocity measurements to identify an apparent prograde stellar stream in the Solar vicinity, which they named Nyx \citep{Necib2019a}. The stars identified as members of Nyx were found to be on orbits that trail the Galactic disk by about 90~\kms, which is at the edge of the velocity distribution of the thick disk. This structure was explored further in \citet{Necib20}, using abundance data from the publicly available RAVE-on \citep{Casey2017} and GALAH DR2 \citep{Buder18} catalogs. 
Based on the orbital information and elemental abundances of the stars, the authors concluded that Nyx must be the remnant of a disrupted dwarf galaxy. 

The key abundance information applied in that work was the magnesium abundance of the stars relative to iron, \mgfe. The evolution of $\alpha$ element abundances (Mg, Si, S, Ca, Ti) with overall metallicity is sensitive to the mass and star formation history of a galaxy, in the sense that 
the \alphafe ratio, which always begins at a super-Solar level at low metallicity, declines as \feh increases, and the start of that decline occurs at lower \feh in lower-mass galaxies \citep{Venn2004,Tolstoy2009}. Seven of the Nyx stars had been observed by the RAVE survey and had abundance results in the RAVE-on catalog, and when \citet{Necib20} compared their \mgfe values to literature abundances for the thick disk and halo, they found that the stars had systematically lower magnesium abundances than would be expected for thick disk stars at the same \feh, which would suggest they formed in a dwarf galaxy, rather than {\em in situ} in the Milky Way \citep[e.g.,][]{Sheffield2012}.

In this work, we re-examine Nyx using high quality abundance data from the Third Data Release of the GALAH Survey \citep[submitted to MNRAS,  arXiv:2011.02505]{Buder2021} 
and the Sixteenth Data Release of the APOGEE Survey \citep{Ahumada2020}. Looking at the abundances of key elements from these catalogs we find that Nyx in fact shows the same abundance patterns as the $\alpha$-rich thick disk of the Milky Way, and that the abundances considered by \citet{Necib20} --- key evidence for an accretion origin --- were likely erroneous or misinterpreted. We therefore conclude that Nyx is probably simply a group of stars at the tail of the thick disk's kinematic distribution.

This paper is structured as follows: Section \ref{sec:data} introduces the surveys we draw data from, Section \ref{sec:abundance} considers the abundance data for Nyx stars, Section \ref{sec:dwarf} compares the abundances in Nyx to those in the thick disk and dwarf galaxies, and in Section \ref{sec:discuss} we discuss the results.

\section{Data}\label{sec:data}
In this work we make use of abundance data from three large stellar spectroscopic surveys: GALAH DR3, APOGEE DR16, and RAVE-on. 
The Third Data Release \citep[DR3;][]{Buder2021} of the GALactic Archaeology with HERMES (GALAH) survey presents stellar parameters, radial velocities, and up to 30 elemental abundances for 588,571 stars, derived from optical spectra at a typical resolution of $R\sim28{,}000$. We use the following criteria to select a reliable data set for this project: (i) GALAH flag $\texttt{flag\_sp}=0$ (no problems noted in the input data, reduction, or analysis); (ii) GALAH flag $\texttt{flag\_fe\_h}=0$ (no problems noted in the iron abundance determination); (iii) GALAH flag $\texttt{flag\_x\_fe}=0$ when considering an individual abundance [\texttt{x}/Fe]; 
(iv) GALAH flag $\texttt{snr\_c3\_iraf}>30$ (an average signal-to-noise in the red camera greater than 30 per pixel).

The Sixteenth Data Release \citep[DR16;][]{Ahumada2020} of the Apache Point Observatory Galactic Evolution Experiment (APOGEE) contains stellar parameters, radial velocities, and abundances of up to 20 elements for more than 430,000 stars, derived from $H$-band infrared spectra at a resolution of $R\sim22{,}500$. We only considered stars with $\texttt{ASPCAPFLAG} = 0$, indicating no problems in the data, reduction, or analysis, and when considering an individual abundance [\texttt{x}/Fe], we required $\texttt{X\_FE\_FLAG}=0$.

The Fifth Data Release \citep[DR5;][]{ravedr5} of The Radial Velocity Experiment (RAVE) covers 457,588 stars, with medium-resolution spectra ($R\sim7{,}500)$ in the region of the Ca triplet ($\sim8600$\AA). In this work we use results from the RAVE-on catalog \citep{Casey2017}, as this was the catalog primarily used by \citet{Necib20}. RAVE-on is a re-analysis of RAVE DR5 spectra using the data-driven ``label transfer'' method of The Cannon \citep{Ness2015}, and provides stellar parameters and abundances of up to seven elements per star (O, Mg, Al, Si, Ca, Fe, and Ni).  Here we required $\texttt{qc}=1$ (indicating that stars meet data quality constraints), and for stars with multiple spectra, we chose the one with the highest signal-to-noise ratio.

For all surveys, the Galactic orbital properties of the stars were calculated in the same manner as used to construct the GALAH DR3 kinematic value-added catalog. This is described in detail in the GALAH DR3 data release paper \citep{Buder2021}. Briefly, we used \textsc{galpy}, with the \textsc{McMillan2017} potential \citep{McMillan2017} and the values $R_\mathrm{GC}=8.21$~kpc and $v_\mathrm{circular}=233.1~\kms$ \citep{Abuter2019}. We set $(U,V,W)_\mathrm{\odot}=(11.1, 15.17, 7.25)~\kms$ in keeping with \citet{Reid2004} and \citet{Schonrich2010}. For APOGEE and RAVE stars, we used the radial velocities reported by each survey, and distances from \citet{Bailer-Jones2018}. For GALAH we used DR3 radial velocities and distances from the age and mass value-added catalog, which primarily incorporated distances found by the Bayesian Stellar Parameters estimator \citep[\textsc{bstep}; described in][]{Sharma2018}, which calculates distance simultaneously with mass, age, and reddening.

\section{Abundances for candidate Nyx stars}\label{sec:abundance}

The primary aim of our work is to compare Nyx in the context of abundance space to the disk and halo of our Galaxy, as well as to nearby dwarf galaxies. \citet{Necib20} provided a catalogue of 232 high confidence members of Nyx found in \gaia DR2. We cross-matched these stars with each of the surveys described in Section \ref{sec:data} using their \gaia DR2 \texttt{source\_id} and found that 18 Nyx stars had results in GALAH DR3, 19 stars were in RAVE-on, and 9 stars were included in APOGEE DR16. There is one Nyx star in common between all three surveys. We used the velocities and orbital properties of the stars to confirm that the Nyx stars in GALAH, APOGEE and RAVE-on are kinematically unbiased relative to the overall Nyx population. The orbital properties we calculate are consistent with those in \citet{Necib20}; namely, these stars are on prograde orbits (i.e., $\jphi>0$) with relatively large orbital energies for disk stars.

\begin{figure}[ht!]
\plotone{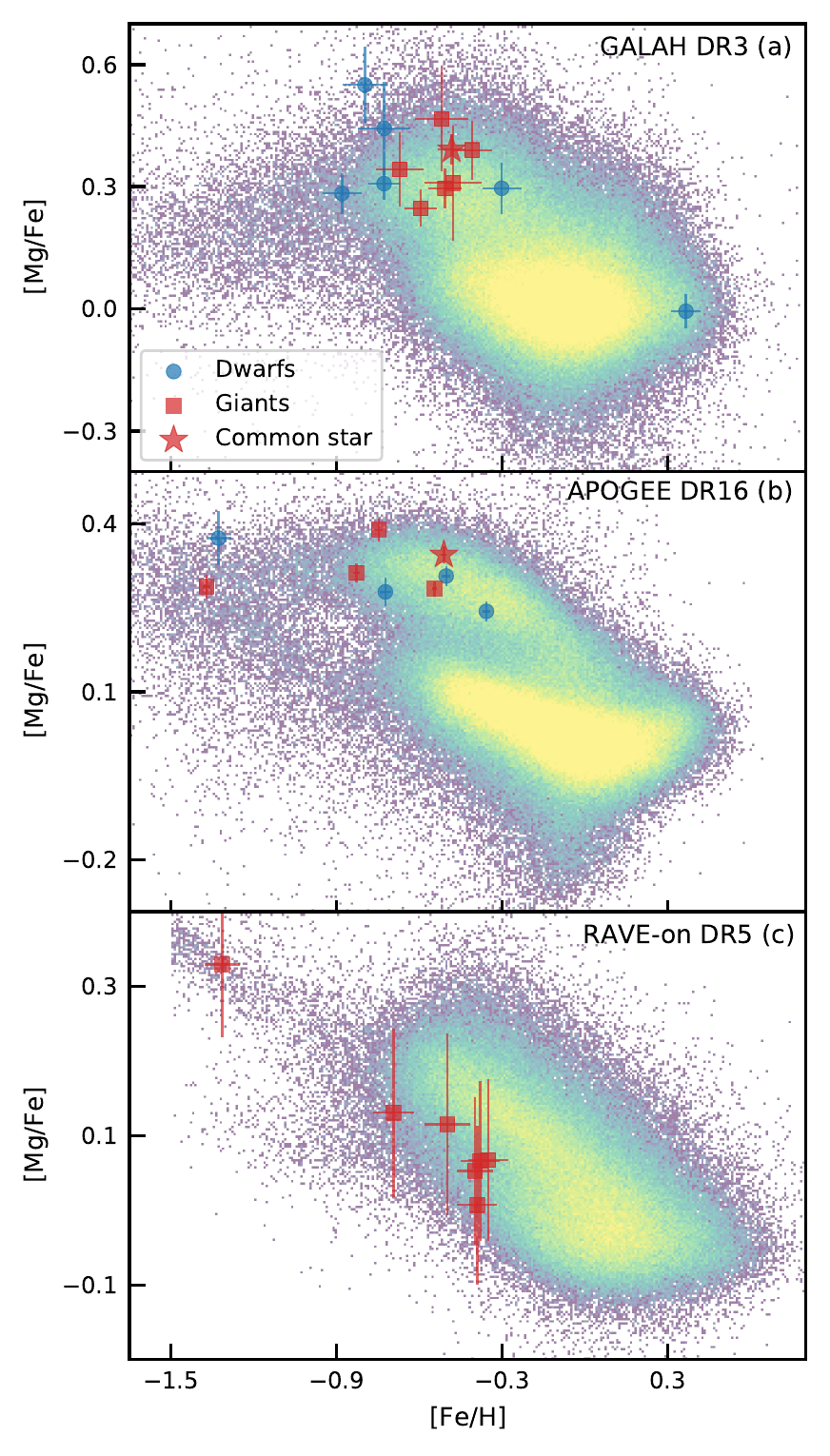}
\caption{Comparison of the \mgfe abundances for Nyx stars identified by \citet{Necib20} as determined by (a) GALAH DR3, (b) APOGEE DR16, and (c) RAVE-on. The background distribution on each panel is a scatter log-density plot of all stars from a given catalogue. For each survey the dwarfs and giants in Nyx have been identified by their \logg ($\logg > 3.5$ or $\logg \leq 3.5$, respectively). We find that GALAH DR3 and APOGEE DR16 place the Nyx stars in the same range of \mgfe as the thick disk, while RAVE-on locates Nyx stars at lower \mgfe values, as presented in \citet{Necib20}. Note that none of the RAVE-on dwarfs (nor the one common star, which is a giant) had \mgfe abundances reported by RAVE-on.
\label{fig:nyx_each_survey}}
\end{figure}

Figure \ref{fig:nyx_each_survey} shows the \feh-\mgfe plane as recovered by each of the three surveys. The background distribution includes all stars that meet the relevant quality flags. Each survey is dominated by the disk of the Milky Way, so we see the canonical $\alpha$-poor and $\alpha$-rich populations. On each panel we highlight the Nyx stars, as identified by \citet{Necib20}, that were observed by each survey. Those stars are split almost evenly between dwarfs and giants (defined as $\logg > 3.5$ and $\logg \leq 3.5$, respectively); however, none of the RAVE-on dwarfs (nor the one common star, which is a giant) had \mgfe abundances measured by RAVE-on. 

The key piece of evidence that \citet{Necib20} used to conclude that the Nyx stars were accreted was the \mgfe abundance data from RAVE-on. They noted that the stars had low \mgfe for their metallicity, compared to thick disk abundances collated from the literature by \citet{Venn2004}. This direct comparison between RAVE-on and literature abundances could be problematic, but from Figure \ref{fig:nyx_each_survey}c, we would draw the same conclusion as \citet{Necib20} --- in the RAVE-on data set, the Nyx stars are systematically lower in \mgfe for their metallicity compared to the $\alpha$-rich thick disk.

However, the sample of Nyx stars in the GALAH DR3 and APOGEE DR16 catalogues sit squarely in the region of the \feh-\mgfe plane occupied by the thick disk. We only show \mgfe here, but this consistency with the thick disk also holds for other $\alpha$ elements. There are a number of potential explanations for this, including the possibility of a systematic issue with RAVE-on $\alphafe$ abundances, or, alternatively, the existence of a large intrinsic $\alphafe$ scatter in Nyx, with the subset of stars with RAVE-on abundances coincidentally all having low \mgfe.

\begin{figure}[ht!]
\plotone{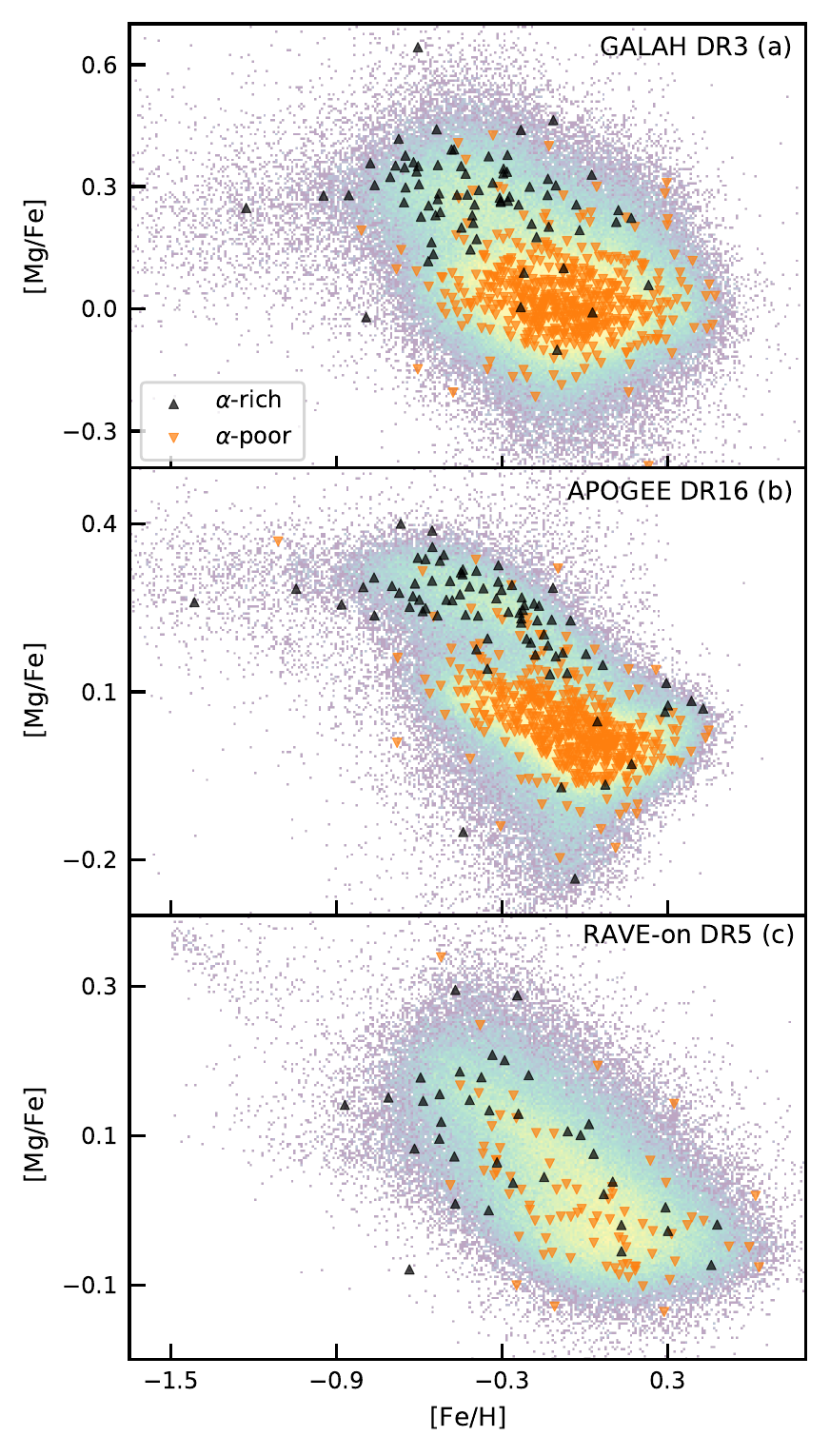}
\caption{There are 891 stars in common between GALAH DR3 (top), APOGEE DR16 (middle) and RAVE-on (bottom). The background distribution in each panel is a scatter log-density plot of all stars with reliable \mgfe from the appropriate catalog. Highlighted are the all stars in common between each survey, categorized by their GALAH \alphafe, with upward black triangles showing stars with $\alphafe> 0.16$, and downward orange triangles having $\alphafe \leq 0.16$. While GALAH and APOGEE \mgfe abundances follow the GALAH \alphafe well, the RAVE-on \mgfe abundances for the two groups have larger scatter and similar mean values.  \label{fig:overlap}}
\end{figure}

Pursuing this first explanation, we look at the 891 stars in common between GALAH DR3, APOGEE DR16, and RAVE-on (as identified by \gaia DR2 \texttt{source\_id}). It should be noted that direct comparison of the surveys requires us to compare results from GALAH and APOGEE, which have broad spectral coverage at high resolution ($R > 20,000$), to RAVE, whose spectra covered only the calcium triplet region at moderate resolution ($R \sim 7,500$). There are also differences in data quality between the surveys due to differing target selection choices and instrumental capabilities. The common stars have data quality that is typical for RAVE, while they are somewhat low in signal to noise for APOGEE.

As we are comparing \mgfe values across three different surveys, it is important to look for systematic differences. For the stars in common between APOGEE and RAVE-on, average values from the former are larger than those from the latter by $\Delta(\mgfe)=0.08\pm0.09$~dex, while for GALAH compared to RAVE-on, the mean difference is $\Delta(\mgfe)=0.11\pm0.12$~dex. Finally, for APOGEE compared to GALAH, the average difference is only $\Delta(\mgfe)=0.01\pm0.10$~dex.

In Figure~\ref{fig:overlap} we show the \mgfe distributions of the 891 common stars as recovered by each of the surveys. These are classified as $\alpha$-rich or $\alpha$-poor based on their GALAH \alphafe ratios, splitting at $\alphafe=0.16$. Since magnesium is an $\alpha$ element, we would expect the \mgfe ratio determined by each survey to correlate well with the overall $\alpha$ abundance. This is true for GALAH and APOGEE, where there is a clear distinction in \mgfe between the $\alpha$-rich and $\alpha$-poor populations. The RAVE-on results show that, while there is a rough correspondence between our \alphafe selection and \mgfe (i.e., the mean \mgfe for the $\alpha$-rich stars is higher than the mean \mgfe for the $\alpha$-poor stars), there is no obvious high/low $\alpha$ distinction, with both the $\alpha$-rich and $\alpha$-poor groups covering the full range in \mgfe.

From this comparison of the abundance values reported by the three surveys we can say that it is more likely that the RAVE-on abundances used in \citet{Necib20} are imprecise than it is that Nyx has a large intrinsic \alphafe range, and that the RAVE-on Nyx stars are all by coincidence at the low end of the distribution. Nyx stars in both the GALAH and APOGEE data sets have high measured \mgfe abundances that correlate well with high \alphafe ratios, and those abundances do not show a large scatter. 

\section{Comparing Nyx to the thick disk and to dwarf galaxies}\label{sec:dwarf}

\begin{figure}[ht!]
\plotone{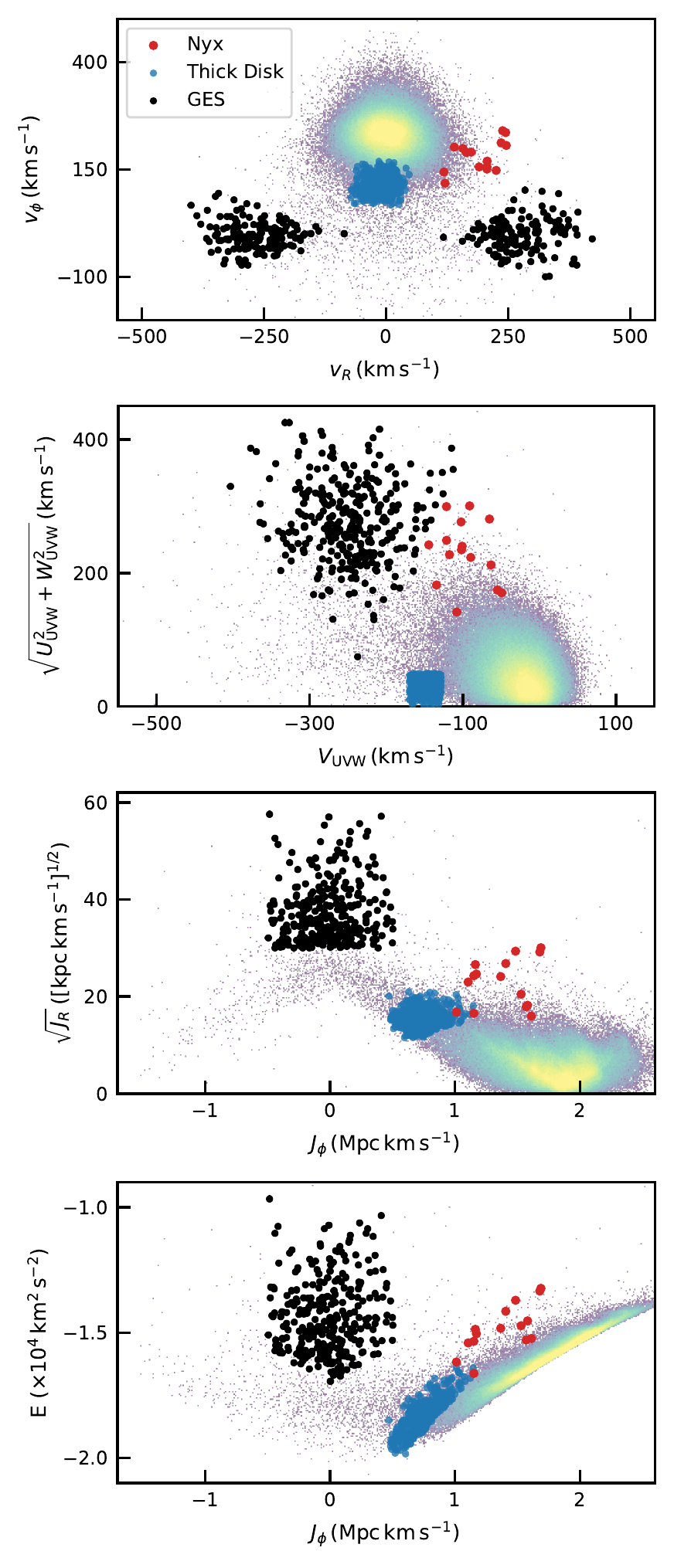}
\caption{The orbital properties of the entire GALAH survey within 3~kpc of the Sun. Highlighted are the Nyx stars (red dots), Gaia-Enceladus stars (black dots), and a kinematic selection of thick disk stars (blue dots); the background distribution in each panel is a scatter log-density plot of all stars.
 \label{fig:galah_nyx_kinematics}}
\end{figure}

\begin{figure}[ht!]
\plotone{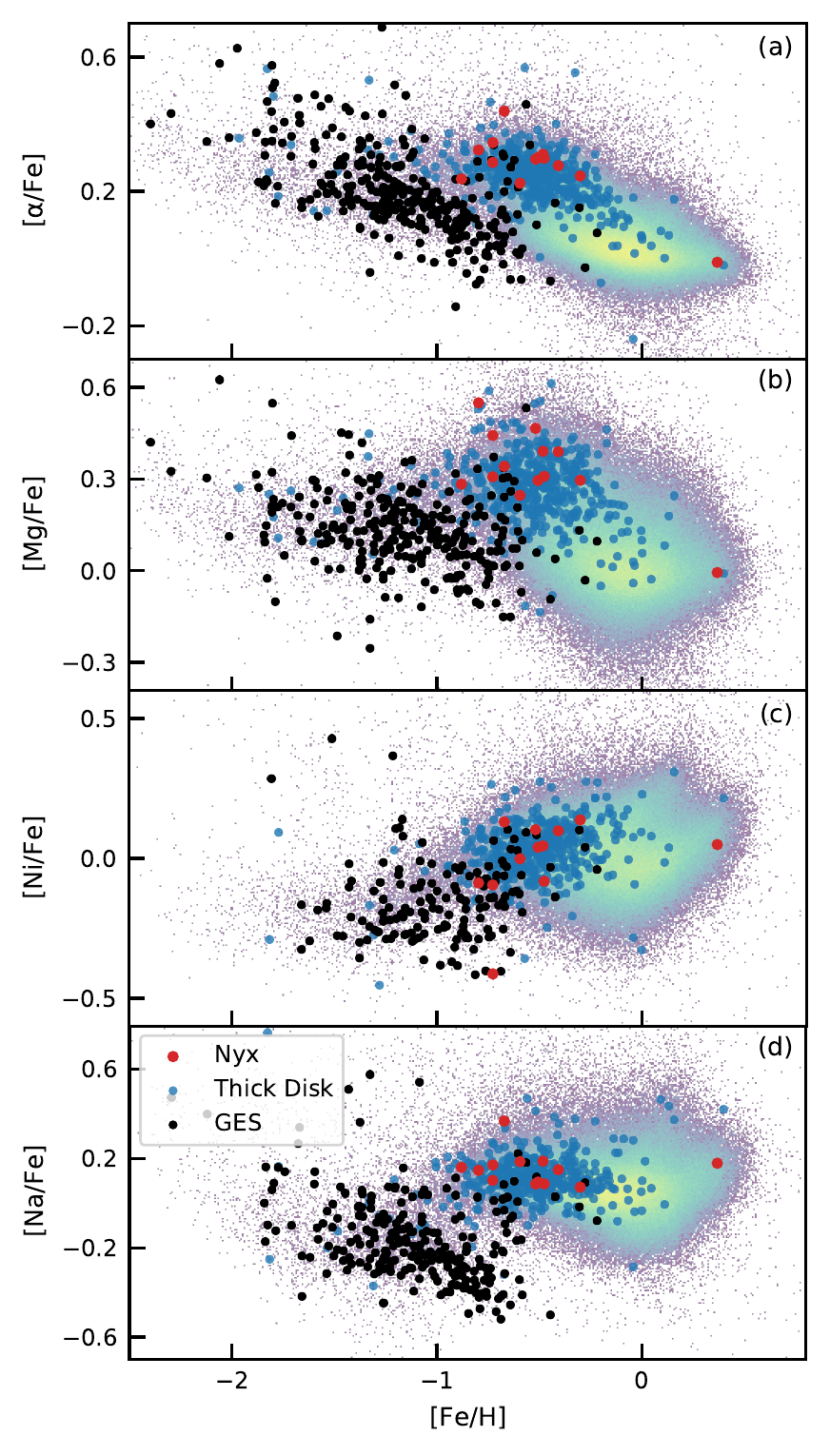}
\caption{Comparison using GALAH DR3 results of Nyx (red dots) to a kinematically-selected sample of thick disk stars (blue dots), and stars from the Gaia-Enceladus accretion event (black dots). \label{fig:dwarf_galaxies_overlap_galah}}
\end{figure}

\begin{figure}[ht!]
\plotone{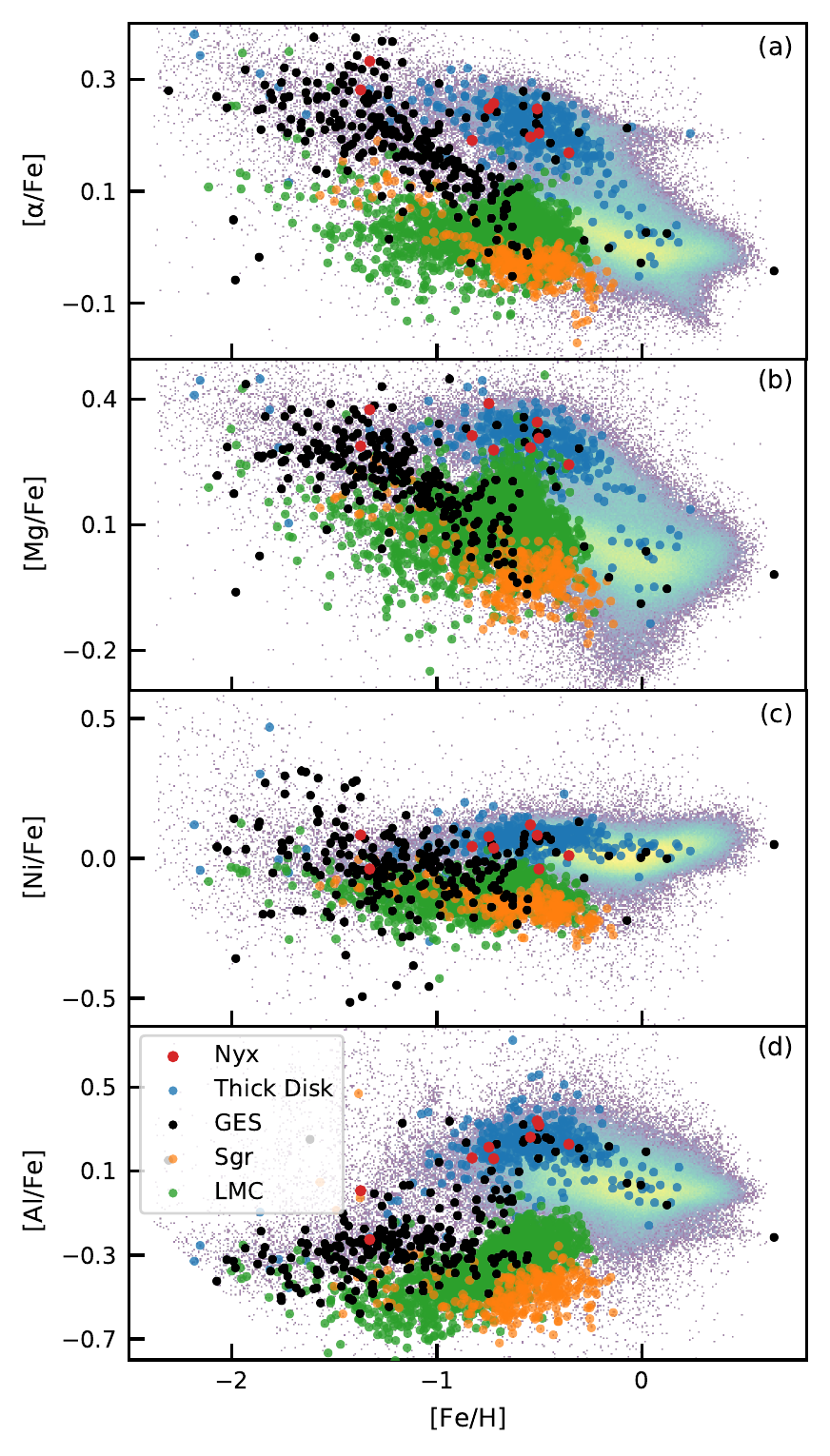}
\caption{Comparison of APOGEE results for Nyx (red dots) to a sample of thick disk stars (blue dots), stars in the Sagittarius dwarf (orange dots), stars in the LMC (green dots) and stars from the Gaia-Enceladus accretion event (black dots). \label{fig:dwarf_disk_comparison_apogee}}
\end{figure}

The principal aim of the GALAH and APOGEE surveys is to measure the abundances of elements from different stellar nucleosynthetic channels. Here we use these abundance sets to consider the similarity between Nyx stars and those from different formation environments.

In particular, we consider the accretion origin hypothesis from \citet{Necib20} by comparing Nyx stars to those from nearby dwarf galaxies. The differing galactic chemical enrichment timelines of dwarf galaxies \cite[e.g.,][]{Venn2004} mean that the stars that form in them show different abundance patterns to those of \textit{in situ} Milky Way stars, e.g., lower [$\alpha$/Fe] abundance ratios than Milky Way stars at a given metallicity. 
In this section we compare the Nyx stars to samples of stars from three dwarf galaxies: the Sagittarius Dwarf, the Large Magellanic Cloud, and the accretion remnant Gaia-Enceladus. 
Noting that in Figure \ref{fig:nyx_each_survey} the Nyx stars look broadly like the thick disk, given their similarly high \alphafe ratios, we also select a sample of thick disk stars using kinematic (rather than abundance) criteria for comparison.

The LMC and Sagittarius stars were identified in APOGEE DR16 by first taking all stars with \texttt{programname} set to ``magclouds'' or ``sgr'' and then applying further selections on radial velocity and proper motion. For Sagittarius we then required RV $>100~\kms$ and proper motions within 0.5~$\mathrm{mas}\,\mathrm{yr}^{-1}$ of $(\mu_\mathrm{RA},\mu_\mathrm{Dec})=(-2.71,-1.37)~\mathrm{mas}\,\mathrm{yr}^{-1}$; for the LMC we required RV $>150~\kms$ and proper motions within 1.0~$\mathrm{mas}\,\mathrm{yr}^{-1}$ of $(\mu_\mathrm{RA},\mu_\mathrm{Dec})=(1.85,0.30)~\mathrm{mas}\,\mathrm{yr}^{-1}$. The Gaia-Enceladus stars were selected from both surveys using angular momentum, \jphi, and radial action, \jr, \citep[as employed by][]{Feuillet2020, Simpson2020a}, requiring $(-0.5<\jphi<0.5)~\jphiunits$ and $\sqrt{\jr}>30~\jrunits$. The \jr-\jphi selection identifies many of the same stars as the different kinematic selections used by other researchers, but it has the advantage of less contamination from non-Gaia-Enceladus stars.  We selected comparison sets of thick disk stars from both GALAH and APOGEE for comparison against Nyx by taking all stars in the region of Toomre space with $|V_{\rm UVW}+150~\kms| < 20~\kms$ and ($0 \leq \sqrt{U_{\rm UVW}^{2} + W_{\rm UVW}^{2}} \leq 50)~\kms$, well away from the Nyx stars.

Figure~\ref{fig:galah_nyx_kinematics} shows the orbital properties of all GALAH stars within 3~kpc of the Sun in four different kinematic coordinate systems. The Nyx stars are shown as red dots, and they clearly do have orbital properties at the extreme end of the distribution of disk stars. The thick disk comparison stars are plotted as blue dots, and the Gaia-Enceladus stars are black dots. This figure shows that the three groups can be more clearly separated in some kinematic spaces than others. In Figure~\ref{fig:dwarf_galaxies_overlap_galah}, we follow the overall GALAH sample and these three specific groups through the \alphafe, \mgfe, \nife, and \alfe abundance planes; 
the latter two elements are chosen because they represent different stellar nucleosynthetic channels. Here we see that the Gaia-Enceladus stars have markedly lower abundances of all these elements at a given metallicity than the thick disk and the Nyx stars. Conversely, the abundances of the Nyx stars are indistinguishable from those of the thick disk stars, although they are on the edge of the orbital parameter distribution. 

Figure~\ref{fig:dwarf_disk_comparison_apogee} makes a similar comparison using APOGEE data, and adds stars from Sagittarius (in orange) and the LMC (in green), as those fall within the larger volume explored by that survey. The stellar abundance patterns of these additional galaxies are broadly similar to the Gaia-Enceladus stars and different from the overall APOGEE data set, the thick disk, and Nyx. Again, Nyx is more similar to the thick disk in these abundance planes than it is to dwarf galaxies. What overlap there is with the dwarf galaxies is confined to the two most metal-poor stars, in the regime where the dwarf galaxies are difficult to distinguish from the Galactic halo, and where stars from Gaia-Enceladus 
appear coincident with the metal-poor end of the thick disk (although it is not clear if this latter feature represents real overlap in abundance space or simply thick disk contamination in the Gaia-Enceladus selection).

\section{Discussion}\label{sec:discuss}
We would expect stars accreted from dwarf galaxies to be distinct from stars formed \textit{in situ} in a number of their properties. They should be clustered in orbital parameter space, and possibly spatially; their age distribution should be truncated at the time of accretion, with no younger stars; and their abundance patterns should show key signs of their low-mass formation environment, with a lower minimum metallicity and faster \alphafe depletion, as well as potentially other chemical signatures \citep[e.g.,][]{Das2020,Nissen2011,Casey2014a,Casey2014b}. In this study we investigated those aspects of the Nyx stream using data from GALAH DR3, APOGEE DR16, and RAVE-on.

We calculated orbital parameters for the Nyx stars in all three spectroscopic data sets, confirming that the Nyx stars are kinematically similar to each other, and have higher orbital energies and eccentricities than typical thick disk stars. 

\citet{Necib20} find that the colour-magnitude diagram of their stars is consistent with an older isochrone. They take this as support for Nyx being a discrete accreted population, as the Milky Way's disk displays a range of ages. However, age is not a strong discriminant in this case because the thick disk has been shown to be relatively old \citep[e.g.,][]{Gilmore1985,Sharma2019,Sharma2020b}, with a mean age of 10~Gyr, which is consistent with the age inferred for Nyx by \citet{Necib20}.

Comparing abundances for Nyx stars from GALAH, APOGEE, and RAVE-on against the thin and thick disk and dwarf galaxies, we find no elemental abundance differences between Nyx stars and thick disk stars -- whether selected kinematically or by \alphafe\ -- in the GALAH DR3 and APOGEE DR16 data sets. While RAVE-on abundances do appear to show a difference, we consider the RAVE-on \mgfe abundances potentially unreliable for these stars, 
as there is no clear distinction between high and low \alphafe\ stars in our Figure \ref{fig:overlap} or in the underlying RAVE dataset \citep{wojno2016}.

Analysis of abundance information from both the GALAH and APOGEE surveys 
shows that members of the LMC, Sagittarius and Gaia-Enceladus can be distinguished rather well from Galactic thin and thick disk populations -- and from each other -- across the space of $\alpha$ elements, iron-peak elements, and light odd-Z elements. In that same abundance space, the kinematically-identified Nyx members are entirely consistent with the thick disk. We note that comparisons of this sort are best done within surveys rather than across surveys, as this tends to minimize the effects of differing systematics in the underlying data and their analysis.

The disk and spiral structure of the Milky Way display large-scale perturbations which have been attributed to close interactions with satellites such as the Sagittarius dwarf \citep[e.g.,][]{Antoja2018,BlandHawthorn2019}, so it is entirely plausible that Nyx could be the {\em result} of an accretion event even if its constituent stars formed {\em in situ} in the Milky Way \citep[see, e.g.,][]{JeanBaptiste2017}.  Previous studies have revealed the existence of a comparatively metal-rich (\feh $> \sim -1$) stellar population with \alphafe abundances typical of the thick disk on prograde halo-like orbits in the solar neighborhood \citep[e.g.,][]{Bonaca2017,Belokurov2020}; the latter work postulated that this population consisted of Galactic proto-disk stars heated by a massive merger event, and noted that Nyx could simply be part of this larger population.  

Hence Nyx likely joins other kinematically identified streams such as Aquarius \citep{Williams2011,Casey2014b} and Hercules \citep{Famaey2005,Bensby2007} in the category of stellar substructures created by past interactions or secular processes, and not the debris of a disrupted satellite.

\acknowledgments
The GALAH survey is based on observations made at the Anglo-Australian Telescope, under programs A/2013B/13, A/2014A/25, A/2015A/19, A/2017A/18, A/2019A/15, A/2020B/23. We acknowledge the traditional owners of the land on which the AAT stands, the Gamilaraay people, and pay our respects to elders past and present. This paper includes data that have been provided by AAO Data Central (\url{datacentral.org.au}).
This work has made use of data from the European Space Agency (ESA) mission {\it Gaia} (\url{https://www.cosmos.esa.int/gaia}), processed by the {\it Gaia} Data Processing and Analysis Consortium (DPAC, \url{https://www.cosmos.esa.int/web/gaia/dpac/consortium}). Funding for the DPAC has been provided by national institutions, in particular the institutions participating in the {\it Gaia} Multilateral Agreement.
Parts of this research were conducted by the Australian Research Council Centre of Excellence for All Sky Astrophysics in 3 Dimensions (ASTRO 3D), through project number CE170100013. DBZ, JDS, and SLM acknowledge the support of the Australian Research Council through Discovery Project grant DP180101791, and ARC acknowleges support through DECRA fellowship DE190100656. SLM and JDS are supported by the UNSW Scientia Fellowship scheme. TZ and JK acknowledge the financial support of the Slovenian Research Agency (research core funding No. P1-0188) and the European Space Agency (PRODEX Experiment Arrangement No. C4000127986). 

\facilities{AAT}
\software{TOPCAT \citep[v4.7-2;][]{Taylor2005},
        numpy \cite[v1.20.1;][]{Harris2020},
        astropy \citep[v4.0.1;][]{astropy18},
        matplotlib \citep[v3.1.3;][]{Hunter2007,Caswell2020}, 
        scipy \citep[v1.4.1;][]{SciPy1.0Contributors2020},
        galpy \citep[v1.6.0;][]{Binney2012,Bovy2013,Bovy2015}}

\bibliography{Nyx}
\bibliographystyle{aasjournal}

\end{CJK*}
\end{document}